\documentclass[9pt,twocolumn,twoside]{osajnl}
\usepackage{physics}
\journal{pr} % Choose journal (ao, aop, josaa, josab, ol, pr)

\setboolean{shortarticle}{false}
\title{Spectrally-resolved Hong-Ou-Mandel interferometry for Quantum-Optical Coherence Tomography}

\author[1]{Pablo Yepiz-Graciano}
\author[1]{Al\'i Michel Angulo Mart\'inez}
\author[2,4]{Dorilian Lopez-Mago}
\author[1]{Hector Cruz-Ramirez}
\author[1,3]{Alfred B. U'Ren}

\affil[1]{Instituto de Ciencias Nucleares, Universidad Nacional Aut\'onoma de M\'exico, Apdo. Postal 70-543, Ciudad de M\'exico 04510, Mexico}
\affil[2]{Tecnologico de Monterrey, Escuela de Ingenier\'ia y Ciencias, Ave. Eugenio Garza Sada 2501, Monterrey, N.L. 64849, Mexico}

\affil[3]{alfred.uren@correo.nucleares.unam.mx}
\affil[4]{dlopezmago@tec.mx}

\begin{abstract}
In this paper, we revisit the well-known Hong-Ou-Mandel (HOM) effect in which two photons, which meet at a beamsplitter, can interfere destructively, leading to null in coincidence counts. In a standard HOM measurement, the coincidence counts across the two output ports of the beamsplitter are monitored as the temporal delay between the two photons prior to the beamsplitter is varied, resulting in the well-known HOM dip. We show, both theoretically and experimentally,  that by leaving the delay fixed at a particular value while relying on spectrally-resolved coincidence photon-counting, we can reconstruct the HOM dip, which would have been obtained through a standard delay-scanning, non-spectrally-resolved HOM measurement.   We show that our numerical reconstruction procedure exhibits a novel dispersion cancellation effects, to all orders. We discuss how our present work can lead to a drastic reduction in the time required to acquire a HOM interferogram,  and specifically discuss how this could be of particular importance for the implementation of efficient quantum-optical coherence tomography devices. 
\end{abstract}

\setboolean{displaycopyright}{true}

\begin{document}

\maketitle

%%%%%%%%%%%%%%%%%%%%%%%%%%%%%%%%%%%%%%%%%%%%%%%%%%%%%%%%%%%%%%%%%%%%%%%%%%%%%%%%%%%%%%%%%%%%%%%%%%%%%%%%%%%%%%%%%%%%%%
\section{Introduction}

The present work lies at the crossroads of experimental quantum optics and biomedical applications. Advances in quantum technologies have made possible an exciting breadth of applications in fields such as communications~\cite{Gisin2007}, imaging~\cite{Moreau2019}, and computation~\cite{Ladd2010}. In this work, we aim to explore the application of quantum optical effects in the field of biomedicine by studying the well-known Hong-Ou-Mandel (HOM)  interference effect in an interesting new light. This effect, through which two photons which meet at a beamsplitter can exhibit quantum interference, represents a hallmark of quantum optics. First demonstrated by Hong \emph{et al}.~\cite{Hong1987}, it relies on the destructive interference which occurs between the reflected-reflected (RR) and transmitted-transmitted (TT) alternatives if these are indistinguishable, leading to null in coincidence counts across the two beamsplitter outputs. It is a remarkable effect that is fundamentally based on the quantum-mechanical nature of the interfering states of light.

In a typical HOM experiment, the signal and idler photon pairs produced by the spontaneous parametric downconversion (SPDC) process reach the two input ports of a beamsplitter with a controllable temporal delay between them. While for sufficiently large delays, the TT and RR alternatives become distinguishable, and the quantum interference is thus inhibited, at zero delay, the well-known suppression of coincidence counts occurs. The resulting coincidences vs. delay curve, which we refer to in this paper as the HOM interferogram, then exhibits a characteristic dip centered at zero delay. The dip characteristics can reveal useful information about the quantum state of the interfering photon pairs. If both interfering photons emanate from a single SPDC source, on the one hand, the observed dip visibility quantifies the degree of symmetry in the photon-pair state upon interchanging the roles of the signal and idler photons. On the other hand, the shape of the dip, as is to be described below, substantially corresponds to the Fourier transform of the photon-pair joint spectral amplitude so that the dip width is inversely proportional to the SPDC bandwidth~\cite{Hong1987}.

An interesting direct consequence of the destructive interference between the TT and RR alternatives, is that the quantum state emanating from the two output ports $\alpha$ and $\beta$ of the beamsplitter exhibits path entanglement, in particular representing the NOON state $2^{-1/2}(\vert 2\rangle_\alpha \vert0 \rangle_\beta \pm \vert 0\rangle_\alpha \vert2 \rangle_\beta)$~\cite{Boto2000}. Such states are attractive because they show quantum-conferred phase super-resolution effect~\cite{Mitchell2004}. Another interesting property of HOM interference is that if the SPDC source used is based on a continuous-wave pump, the HOM visibility is insensitive to even-order dispersion experienced by one or both photons prior to reaching the beamsplitter \cite{Steinberg1992,Franson1992}. This remarkable property led to the proposal of a quantum version of optical coherence tomography (OCT)~\cite{Huang1991}, which is based on a HOM interferometer except that one of the SPDC photons is reflected from a sample under study before reaching the beamsplitter \cite{Abouraddy2002}. Under appropriate conditions, a distinct HOM dip can appear for each interface within the sample, thus yielding useful morphological information about such a sample. Interestingly, because the resolution depends on the HOM dip width, this scheme benefits from the dispersion cancellation effect mentioned above: the instrument's resolution is not affected by even-order dispersion in the sample. In addition, it was shown that this quantum OCT (QOCT) scheme leads to a quantum-conferred factor of $2$ enhancement in resolution as compared to an equivalent classical system with the same bandwidth~\cite{Nasr2003}.

Current advances in photodetection technologies and the development of highly efficient SPDC sources have revived the idea of developing efficient QOCT systems~\cite{Teich2012,Berchera2019,Taylor2016}. A Michelson version of QOCT has been shown to represent a  functional and robust configuration, which can benefit from the incorporation of novel photon-number-resolving detectors~\cite{Micuda2008}, and which is amenable to miniaturization~\cite{Lopez-Mago2012e,Lopez-Mago2012b}. A recent full-field QOCT implementation has been demonstrated using an intensified CCD camera~\cite{Ibarra2020}. This method, akin to full-field OCT, reduces considerably the time required for probing a three-dimensional object by eliminating the need for raster scanning the transverse section of the sample. We have developed a fiber-based QOCT system that incorporates spectrally engineered photon pairs in the telecom regime. We demonstrated interesting interference effects which depend on the type of frequency entanglement~\cite{Yepiz2019}. In terms of axial resolution, beyond the quantum-conferred improvement factor of 2, it is possible to incorporate spectral engineering in the form of chirped, aperiodically poled nonlinear crystals to obtain submicrometer resolutions~\cite{Carrasco2004,Okano2016}; alternatively,  a Fisher information analysis can result in attosecond resolutions~\cite{Lyons2018}. Moreover, it is worth mentioning that another optical sectioning technique employing nonclassical light has recently been demonstrated, called induced coherence tomography~\cite{Valles2018}, which relies on the concept of induced coherence between two downconverters to infer the internal structure of the sample~\cite{Zou1991,Chekhova2016}. 

%It offers the advantage of probing the sample with infrared photons (ideal for biological tissues) while the wavelength of the detected photons is optimum for the detection system.  

While in a typical HOM experiment, the signal and idler photon pairs are detected in a non-spectrally-resolved manner, in the present work we explore the benefits of lifting this restriction and permitting spectrally-resolved coincidence photon counting of the optical modes corresponding to the beamsplitter output ports. Other groups have carried out related experiments in which the joint spectral intensity of the output quantum state is analyzed as the delay is varied~\cite{Gerrits15}. In this paper, we show both theoretically and experimentally that leaving the delay fixed while enabling spectrally-resolved coincidence counting it becomes possible to recover the HOM interferogram, which would have been obtained through a standard delay-scanning, non-spectrally-resolved HOM measurement. We show that this technique permits the recovery of information about the two-photon state with dispersion canceled to all orders. Importantly, we show that from a single delay value larger than the dip half-width, we are able to extract the HOM dip with the same level of background counts as obtained in the standard measurement based on a sufficiently large number of delay stops for the adequate sampling of the dip structure. The importance of this is that the time required to obtain the HOM interferogram can be drastically reduced. As we discuss below, this fixed-delay HOM scheme can be particularly useful in the context of QOCT, for which the standard delay-scanning approach leads to the need for long acquisition times, which is impractical for real-life conditions, e.g., clinical settings. We hope that this work will help to pave the way towards the deployment of QOCT  as a practical technology.

%%%%%%%%%%%%%%%%%%%%%%%%%%%%%%%%%%%%%%%%%%%%%%%%%%%%%%%%%%%%%%%%%%%%%%%%%%%%%%%%%%%%%%%%%%%%%%%%%%%%%%%%%%%%%%%%%%%%%%%%%%
\section{Theory}

The two-photon state produced by SPDC for a continuous-wave (CW) pump, may be written as 
\begin{equation}
\ket{\psi} = \ket{0}_s\ket{0}_i+ \eta \int_{-\infty}^{\infty} \mathrm{d}\Omega  f(\Omega) \ket{\omega_0+\Omega}_s\ket{\omega_0-\Omega}_i, \label{eq:estado_mono}
\end{equation}
where $\omega_0=\omega_p/2$ is the degenerate SPDC frequency, in terms of the pump frequency $\omega_p$, assumed to exhibit a negligible frequency spread.  This state involves the energy-conserving signal $\omega_s=\omega_0+\Omega/2$ and idler $\omega_i=\omega_0-\Omega/2$ frequencies, in terms of a frequency non-degenerate variable $\Omega=\omega_s-\omega_i$. Here, $\eta$ is a constant related to the conversion efficiency and $f(\Omega)$ represents the joint amplitude function given by 
\begin{equation}
f(\Omega)= f_{0}\, \mbox{sinc}\left[ \frac{L}{2} \Delta k(\Omega) \right]\mbox{exp}\left[ i \frac{L}{2} \Delta k(\Omega) \right] F_f(\Omega),
\end{equation}
where $L$ is the crystal length, $F_f(\Omega)$ describes an interference filter acting on the signal and idler photons, and $\Delta k=k_p-k_s(\omega_0+\Omega)-k_i(\omega_0-\Omega)-2 \pi /\Lambda$ is the phase-matching function in terms of the pump $k_p$, signal $k_s$ and idler $k_i$ wavenumbers and the poling period $\Lambda$ (in the case of a periodically poled nonlinear crystal).  $f_{0}$ is a normalization factor, defined so that the integral of $\vert f(\Omega)\vert^2$ over all $\Omega$ yields unity.

A standard HOM interference experiment involves the signal and idler photons from an SPDC source being directed into the two input ports of a beamsplitter. The coincidence count rate across the two output ports of the beamsplitter is monitored as a function of the signal-idler delay $\tau$ introduced prior to the beamsplitter. The result is the well-known HOM dip, exhibiting a null in coincidence counts centered at $\tau=0$ resulting from destructive interference between indistinguishable RR and TT pathways~\cite{Hong1987}. We will refer to the number of coincidence counts in the dip background (i.e., far from the coincidence null) as $R_0$. It has been shown that the standard HOM interferogram can be expressed in terms of $f(\Omega)$ and $\tau$ as follows~\cite{Hong1987}:
\begin{equation}\label{eq:Rc_tau}
 R_{c}(\tau)=\frac{R_0}{2}\int_{-\infty}^{\infty} \mathrm{d}\Omega \left \vert f(\Omega)-f(-\Omega) e^{i\Omega \tau} \right \vert^2.
\end{equation}

Note that in the above expression, the level of background counts $R_0$ depends essentially on the flux of the SPDC photon-pair source. Note also that the frequency integral in Eq.~(\ref{eq:Rc_tau}) corresponds to an experimental situation in which we do not frequency resolve the photons emanating from the HOM beamsplitter output ports. In this paper, we are interested in studying the effect of spectrally resolving these output photons, which corresponds to removing the mentioned frequency integral, thus obtaining a frequency-delay ($\Omega,\tau$) dependent and normalized interferogram of the form
\begin{equation}\label{eq:rc_tau_omega}
r_c(\tau,\Omega)=\frac{1}{2}  \left \vert f(\Omega)-f(-\Omega) e^{i\Omega \tau} \right \vert^2,
\end{equation}
where, evidently, the following relationship between $R_c(\tau)$ and $r_c(\tau,\Omega)$ is obeyed:
\begin{equation}\label{eq:rc_trace_in_omega}
R_c(\tau)=R_0\int_{-\infty}^{\infty} \mathrm{d}\Omega\, r_c(\tau,\Omega).
\end{equation}
It is straightforward to expand Eq.~(\ref{eq:rc_tau_omega}) to obtain the expression
\begin{equation}\label{eq:rc_HOM_develoment}
r_c(\tau,\Omega)=\frac{1}{2} \left \{ A(\Omega)+B(\Omega)e^{-i\Omega\tau}+B^*(\Omega)e^{i\Omega\tau} \right  \},
\end{equation}
in terms of a frequency-symmetrized, real-valued joint spectral intensity $A(\Omega)$, and a complex-valued cross term $B(\Omega)$, i.e.,
\begin{eqnarray}
A(\Omega) &=& \left \vert f(\Omega)\right \vert^2+\left \vert f(-\Omega) \right \vert^2, \\
B(\Omega) &=&-f(\Omega)f^*(-\Omega).
\end{eqnarray}
Clearly, the function $r_c(\tau,\Omega)$ has a delay-independent contribution characterized by $A(\Omega)$, as well as a delay-dependent factor described by $B(\Omega)$. 

\begin{figure}[h!]
\centering
\includegraphics[width=8 cm]{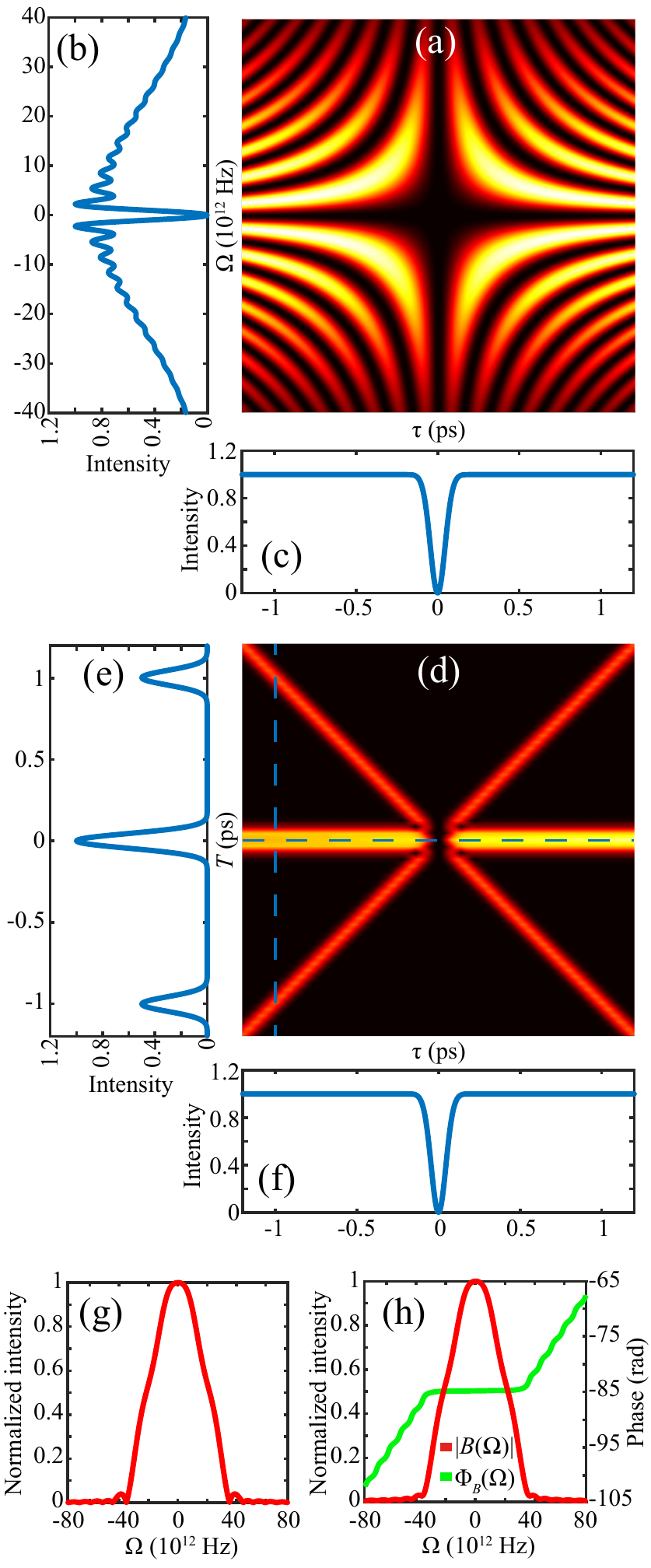}
\caption{(a) Simulation of frequency-delay interferogram $r_c(\tau,\Omega)$. (b) Result of integrating the interferogram over $\tau$, yielding a HOM-like dip in the frequency variable $\Omega$. (c) Result of integrating the interferogram over $\Omega$, yielding the HOM interferogram. (d) Fourier transform of (a), so as to yield the time-domain interferogram $\tilde{r}(\tau,T)$. (e) Evaluation of $\tilde{r}(\tau,T)$ at $\tau=1$~ps.  (f) Evaluation of $\tilde{r}(\tau,T)$ at  $T=0$, yielding the HOM interferogram. (g) Function $A(\Omega)$. (h) Function $B(\Omega)$.
 \label{fig:hom_espec_calc}}
\end{figure}

In order to illustrate these relationships, we present in Fig.~\ref{fig:hom_espec_calc} simulations for a specific experimental situation involving a periodically-poled lithium niobate (PPLN) crystal pumped by a Ti:Sapphire laser operating at $775$ nm in CW mode with poling period $\Lambda=19.1$ $\mu$m. Fig.~\ref{fig:hom_espec_calc}(a) shows a simulation of the delay-frequency interferogram for this situation. As is apparent from the figure, the HOM interferogram exhibits an interesting added richness when the output modes from the beamsplitter are frequency-resolved. Note that at $\tau=0$, there is a null in coincidence counts along with all $\Omega$ values. As the value of $\vert \tau \vert$ increases, oscillations along $\Omega$ appear, with a linearly decreasing period, proportional to $1/\vert\tau\vert$. The function $f(\Omega)$ limits the maximum spread of the frequency-delay HOM interferogram along with the frequency variable $\Omega$.

As is clear from Eq.~(\ref{eq:rc_trace_in_omega}), integrating this frequency-delay interferogram $r_c(\tau,\Omega)$ over $\Omega$ yields the standard HOM interferogram, as shown for the particular situation of the previous paragraph in Fig.~\ref{fig:hom_espec_calc}(c). An interesting possibility is to integrate $r_c(\tau,\Omega)$ over the delay variable $\tau$ instead; we have shown the resulting trace in Fig.\ref{fig:hom_espec_calc}(b). Physically, this would represent the effect of averaging over all temporal delays, while monitoring the coincidence rate vs. the frequency variable  $\Omega$. It is interesting that a HOM-like dip also appears vs. the frequency variable as can be seen in Fig.~\ref{fig:hom_espec_calc}(b), with the interpretation that the frequency-degenerate pairs (with $\Omega=0$) lead to indistinguishable RR and TT pathways with the effect of destructive interference and null in coincidence counts. In fact, this result shows a nonlocal interference effect, where destructive interference occurs even when the photons never meet at the beam splitter~\cite{Pittman1996}. 

For the specific case shown in Fig.~\ref{fig:hom_espec_calc}(a) we also show, in Fig.~\ref{fig:hom_espec_calc}(g), a plot of the function $A(\Omega)$  which by construction is symmetric in $\Omega$, and in Fig.~\ref{fig:hom_espec_calc}(h) plots of $\vert B(\Omega) \vert$ and $\Phi_{B}=\mbox{arg}(B(\Omega))$. These functions will play an important role below. In order to continue with our discussion, it is helpful to Fourier transform the interferogram $r_c(\tau,\Omega)$ in the $\Omega$ variable, thus obtaining a 2D  time-domain interferogram as follows:
\begin{equation}
\tilde{r}_c(\tau,T)=\frac{1}{2 \pi} \int_{-\infty}^{\infty} \mathrm{d} \Omega \, r_{c}(\tau,\Omega) e^{i T \Omega},
\end{equation}
in terms of $T$, the Fourier conjugate variable to $\Omega$. Figure~\ref{fig:hom_espec_calc}(d) shows a plot of the function $\tilde{r}_c(\tau,T)$ corresponding to the same situation as in Fig.~\ref{fig:hom_espec_calc}(a). It is interesting to point out that the standard HOM interferogram $R_c(\tau)$ can be obtained by evaluating the 2D time-domain interferogram at $T=0$, i.e., $R_c(\tau)=\tilde{r}_c(\tau,0)$. Furthermore, it is straightforward to show that $\tilde{r}_c(\tau,T)$ can be expressed in terms of $\tilde{A}(T)$, $\tilde{B}(T)$, and $\tilde{B}^{*}(T)$, which correspond to the Fourier transform of the functions $A(\Omega)$,  $B(\Omega)$, and  $B^{\ast}(\Omega)$, respectively, as
\begin{eqnarray}\label{eq:rc_tau_t}
\tilde{r}_c(\tau,T)= \frac{1}{2}\left\{\tilde{A}(T) + \tilde{B}(T- \tau)+\tilde{B}^{\ast} (T+\tau)\right\}.
\end{eqnarray}

%&=&\tilde{A}(T)-\frac{1}{2\pi} \int d \Omega B(\Omega) e^{i(T-2 \tau)\Omega}-\frac{1}{2\pi} \int d \Omega B(\Omega) e^{i(T+2 \tau)\Omega} \\ 

%\begin{figure}[htbp]
%\centering
%\includegraphics[width=8cm]{hom_espec_calc_fft.pdf}
%\caption{(Color online) Espectrograma de HOM calculado. \label{fig:hom_espec_calc}}
%\end{figure}

Note that the HOM interferogram in any of its forms, i.e., $r_c(\tau,\Omega)$, $\tilde{r}_c(\tau,T)$, or $R_c(\tau)$, is fully determined by the two functions $A(\Omega)$ and $B(\Omega)$. While a standard HOM interferogram $R_c(\tau)$ is obtained through a non-frequency-resolved coincidence measurement across the two HOM beamsplitter output modes as a function of the delay $\tau$, we show below that from a \emph{frequency-resolved HOM interferogram at a fixed delay}, i.e., $r_c(\tau_0,\Omega)$ for $\tau=\tau_0$, we can extract functions $A(\Omega)$ and $B(\Omega)$, and subsequently numerically compute the HOM interferogram $R_c(\tau)$, which would have been obtained through a delay-scanning measurement. Importantly, we will show that frequency-resolved HOM data for a single delay value $\tau_0$, selected to lie outside of the standard HOM dip so that we have access to the full background coincidence counts $R_0$, contains the same information as a standard delay-based HOM measurement, assuming that the acquisition time per data point is the same in both measurements (a single point for the frequency-resolved measurement vs. a collection of points for the standard measurement).

Indeed, for a sufficiently large fixed value of the delay $\tau_0$, the resulting time-domain interferogram  $\tilde{r}_c(\tau_0,T)$ yields three distinct peaks corresponding to each of the terms in Eq.~(\ref{eq:rc_tau_t}). By sufficiently large, we mean that the fixed delay $\tau_0$ must be larger than the temporal width of each of the three peaks, so that these do not overlap - note that because the peak widths are related to the HOM dip width, this translates into setting the delay $\tau_0$ to lie outside of the dip. It is notable that from an experimental measurement at a fixed delay $\tau_0$, it then becomes possible to extract functions $A(\Omega)$ and $B(\Omega)$ through the following series of steps: 1) select a value of delay $\tau_0$ and experimentally obtain the function $r_c(\tau_0, \Omega)$ through a spectrally-resolved Hong-Ou-Mandel interferometer, 2) numerically compute Fourier transform, thus obtaining function $\tilde{r}_c(\tau_0,T)$, 3) numerically filter each of the three peaks in turn, 4) compute a numerical inverse Fourier transform for each of the three peaks, 5) multiply the result by the delay-dependent phases $1$, $\exp(i\tau_0\Omega)$, and $\exp(-i\tau_0\Omega)$ (resulting from the Fourier shift theorem for each of the three peaks, respectively) and complex conjugate the data from the right-hand side-peaks, so as to obtain the function $A(\Omega)$ in the case of the central peak and the function $B(\Omega)$ from any of the two side-peaks. In fact, the two side-peaks contain duplicate information, and it is thus only necessary to carry out this procedure for the central peak and one of the side-peaks. However, in order to take advantage of all coincidence counts $R_0$ (distributed among the three peaks) thus ensuring the best possible reconstruction for a given level of counts, it is helpful to estimate function $B(\Omega)$ as the sum of the two side-peaks recovered from the procedure above, divided by two.

The functions $A(\Omega)$ and $B(\Omega)$ obtained in the manner described in the previous paragraph can then be substituted into Eqs.~(\ref{eq:rc_HOM_develoment}), and (\ref{eq:rc_trace_in_omega}), so as to numerically compute the standard HOM interferogram $R_c(\tau)$. Because all $R_0$ coincidence counts appearing in the standard HOM dip background are employed in the $R_c(\tau)$ interferogram reconstruction, and because we utilize  Eqs.~(\ref{eq:rc_HOM_develoment}), and (\ref{eq:rc_trace_in_omega}) (which model the standard HOM effect) in order to predict the interferogram $R_c(\tau)$ at a fixed delay $\tau_0$, the reconstructed and directly-obtained interferogram are in fact expected to be equivalent. It is remarkable that at a fixed delay $\tau_0$ in the dip background, i.e. at a delay location exhibiting a flat dependence on $\tau$ and therefore no useful information, enabling a frequency-resolved HOM measurement permits the full extraction of the standard HOM dip.

In contrast to a standard HOM measurement, a frequency-resolved HOM measurement importantly permits a delay-dependent separation of the three contributions in Eq.~(\ref{eq:rc_tau_t}), which is the basis of our reconstruction protocol. In addition, we note that function $A(\Omega)$ does not depend on the phase of the joint spectral function $f(\Omega)$ (while the function $B(\Omega)$ is phase-dependent), leading to the important additional implication that this delay-enabled separation of the three terms allows the reconstruction of the symmetrized joint spectrum $\vert f(\Omega) \vert ^2 +\vert f(-\Omega) \vert^2$ without dispersive effects. Note that in the case where the joint spectrum is symmetric in the sense that $f(\Omega)=f(-\Omega)$, the functions $A(\Omega)$ and $B(\Omega)$ become identical, aside from a factor of $-2$; in this case, remarkably, our technique permits the reconstruction of the standard HOM interferogram with full dispersion cancellation. This is an interesting addition to a number of dispersion cancellation effects already studied~\cite{Abouraddy2002,Nasr2003,Donnell2011,Okano2013}. However, note that while dispersion suppression, as studied in Refs. \cite{Donnell2011} and \cite{Okano2013}, occurs only for even-order dispersion terms, our numerical method removes all dispersion effects from the symmetrized joint spectrum and/or the HOM dip, depending on the symmetry characteristics of the function $f(\Omega)$.

Let us now address the question of how the resolution of the apparatus used for frequency-resolving the photons emanating from the HOM beamsplitter output ports affects the performance of our HOM reconstruction protocol. We know from the Nyquist sampling theorem that for a band-limited function (with maximum frequency component $\varphi_{s}$ in its spectrum) a sampling period bounded by $\pi / \varphi_{s}$, is sufficient to fully reconstruct the function in question. In our case, the function we wish to determine is $r_c(\Omega,\tau_0)$ and its spectrum is $\vert\tilde{r}_c(T,\tau_0) \vert^2$. Therefore, $r_c(\Omega,\tau_0)$  will be appropriately sampled by a sampling period $\pi/(\vert\tau_0\vert+\delta t)$ where $\vert\tau_0\vert$ is the location of the side peak with half-width $\delta t$. Disregarding the half-width, i.e., $\vert\tau_0\vert + \delta t \rightarrow \vert\tau_0\vert$, and letting $\delta \omega$ be the minimum resolvable frequency interval in our apparatus, we arrive at the conclusion that our protocol is able to reconstruct the HOM dip for delays which fulfill
\begin{equation}
\vert \tau \vert < \frac{\pi}{\delta \omega}.
\end{equation}
Clearly, as the frequency resolution of the apparatus is improved  (i.e., as $\delta \omega$ is reduced), we are able to reconstruct a HOM interferogram over a longer stretch of delay values $\tau$.

%%%%%%%%%%%%%%%%%%%%%%%%%%%%%%%%%%%%%%%%%%%%%%%%%%%%%%%%%%%%%%%%%%%%%%%%%%%%%%%%%%%%%%%%%%%%%%%%%%%%%%%%%%%%%%%%%%%%%%%%%%%%%%

\section{Quantum-optical coherence tomography}

One of the natural applications for HOM interferometry is QOCT. A QOCT apparatus is closely based on a HOM interferometer, except that one of the SPDC photons is reflected from a sample under study, instead of from a mirror, before reaching the beamsplitter. As is well known, each interface in the sample will produce a HOM dip, along with a cross-interference structure (dip or peak) for each pair of surfaces~\cite{Yepiz2019}. In principle, it becomes possible to determine the internal morphology of the sample (number and position of interfaces) from the resulting QOCT interferogram.

Let us consider a hypothetical sample of thickness $L$ and index of refraction $n$. The QOCT interferogram will include a HOM dip corresponding to each of the ends of the sample, along with additional dips for possible intermediate interfaces. If the temporal delay for the HOM interferometer is introduced by a displaceable mirror, the two end dips will be separated by a displacement $2 n L$ of this mirror. Note that the QOCT resolution is determined by the HOM dip width, which is, in turn, inversely proportional to the SPDC anti-diagonal bandwidth. When carrying out an experimental run, one needs to displace the mirror over the distance $2 n L$ with sufficiently small steps so as to be able to determine the location of any possible dips associated with additional interfaces within this range. If the dip width is $\Delta \tau$, or $c \Delta \tau$ expressed as the required mirror displacement, and if we necessitate $M$ points within each dip so as to correctly determine its position, we need a total number of delay stops $N$ in the experiment given by
\begin{equation}
N=\frac{2 L n M}{c \Delta \tau}. \label{eq:Ndelaystops}
\end{equation}

\begin{figure}[h!]
\centering
\includegraphics[width=8cm]{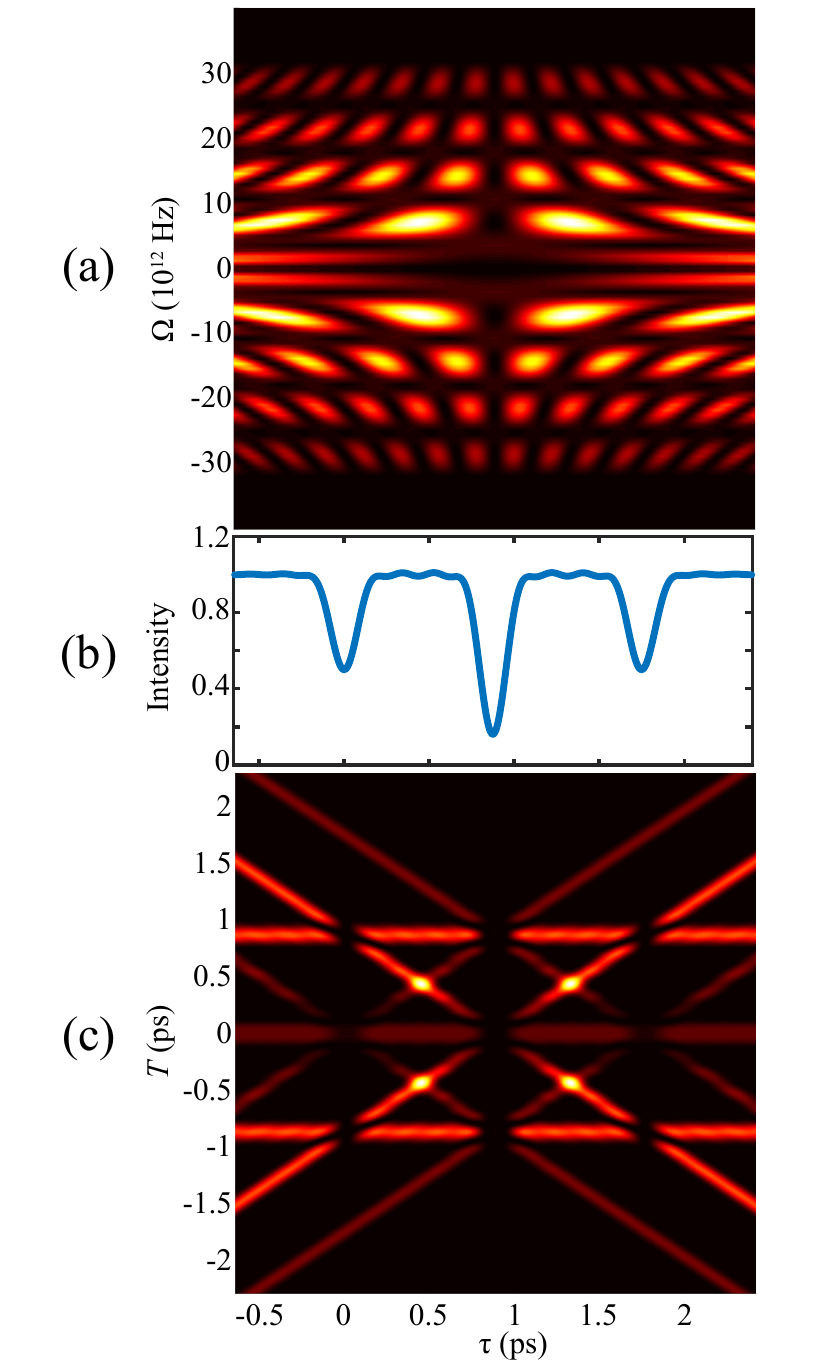}
\caption{(a) Frequency-delay interferogram $r_c(\tau,\Omega)$ for two-interface sample (borosilicate coverslip of $170$~$\mu$m thickness). (b) Result of integrating the interferogram over $\Omega$, yielding the HOM interferogram. (c) Fourier transform of (a) yielding the time-domain interferogram $\tilde{r}_c(\tau,T)$.
 \label{fig:qoct_espec_calc}}
\end{figure}

As an example, for a sample of thickness $L=1$ mm and index of refraction $n=1.5$, with a HOM dip width of $c \Delta \tau=3$ $\mu$m and assuming that $M=3$ points per dip are required for the correct identification of all dips, leads to the need for $3000$ delay stops (displaceable mirror positions). Assuming that the source brightness is large enough so that an acquisition time of $1$ s per point is sufficient, this translates into a total measurement time of 50 minutes (disregarding the time it takes the motor to move from one position to the next). This leads us to discuss one of the essential challenges for the application of QOCT in practical situations: useful data for an unknown sample often requires experimental runs of long duration, which may be impractical in real-life situations, e.g., in a clinical setting in which the sample could be a human eye. This also leads us to discuss one of the key advantages of our protocol for the reconstruction of a HOM interferogram: for a sufficient frequency resolution (which as was discussed above determines the maximum delay $\vert \tau \vert$ which can be recovered in our reconstruction), comparable data to the standard HOM measurement requiring $N$ delay stops (see Eq.~(\ref{eq:Ndelaystops})) can be obtained with a single delay stop (with the same acquisition time per point). In other words, the reduction factor in the required time for an experimental run can be in the thousands, making this technology potentially much more suitable for real-life situations, including clinical settings.

While discussing the applicability of our work, evidently, three-dimensional sample reconstruction is likely to be needed in most real-life situations. In a recent paper from our group~\cite{Ibarra2020}, we have demonstrated full-field QOCT (in which we recover the transverse as well the axial sample structure) by using a triggered intensified CCD camera to detect one of the optical modes following the HOM beamsplitter. It is conceivable to combine spatially- and spectrally-resolved single-photon detection so as to render our present technique full-field capable.

In terms of the application of our current work to QOCT, let us discuss a two-surface sample, which could be regarded as the simplest possible sample of interest for proof-of-principle purposes. In this case, besides the temporal delay $\tau$ and the variable $T$ (Fourier conjugate to $\Omega$), there is a third temporal variable of interest, the optical thickness of the sample in temporal units, $T_s=2 n L/c$. For simplicity, let us assume that the joint spectrum $f(\Omega)$ is symmetric in the sense that $f(-\Omega)= f(\Omega)$, and let us define a new function $F(\Omega)= \vert f(\Omega)\vert ^2$, along with its Fourier transform $\tilde{F}(T)$. Note that if the joint spectrum is not symmetric, it can be rendered symmetric with an appropriate bandpass filter. We can then show that the time-domain interferogram $\tilde{r}_c(\tau,T)$ can be expressed as 
\begin{align}
    \tilde{r}_c = 2 \tilde{F}(T)&+\frac{1}{2}\tilde{F}(T-\tau)+\cos(\omega_0 T_s)\tilde{F}(T-T_s/2)\nonumber \\ 
    & +\cos(\omega_0 T_s)\tilde{F}(T-(\tau+T_s/2))+\frac{1}{2}\tilde{F}(T-(\tau+T_s)) \nonumber \\
    &+\frac{1}{2}\tilde{F}(T+\tau)+\cos(\omega_0 T_s)\tilde{F}(T+T_s/2) \nonumber \\
    &+\cos(\omega_0 T_s)\tilde{F}(T+(\tau+T_s/2))+\frac{1}{2}\tilde{F}(T+(\tau+T_s)).\label{eq:espectrograma_qoct_aprox}
\end{align}

In Fig.~\ref{fig:qoct_espec_calc}(a) we present a plot of the frequency-delay interferogram $r_c(\tau,\Omega)$ expected for a two-layer sample which consists of a borosilicate glass coverslip of $170$ $\mu$m thickness. In panel (b) of this figure, we show the QOCT interferogram obtained by integrating $r_c(\tau,\Omega)$ over  $\Omega$, exhibiting two HOM dips on the sides, each related to one of the two interfaces, as well the corresponding cross-interference intermediate structure in the center. Note that as is well known, e.g., see Ref.~\cite{Yepiz2019}, such a cross-interference intermediate structure will appear in the QOCT interferogram for each pair of surfaces in the sample. In panel (c) we show a plot of the time-domain interferogram $\tilde{r}_c(\tau,T)$, showing for a fixed delay up to $9$ peaks, as is expected from Eq.~(\ref{eq:espectrograma_qoct_aprox}).

Let us note that the placement of these $9$ peaks is symmetric, i.e., there is a central peak and each peak at $T>0$ has a corresponding identical peak at $-T$, so that we need only concern ourselves with the central peak along with the four right-hand-side peaks (which appear on the first two lines of Eq.~(\ref{eq:espectrograma_qoct_aprox})). Note that the first, third and seventh terms in the equation are delay-independent, leading to the appearance of three horizontal stripes in Fig.~\ref{fig:qoct_espec_calc}(c) (the central one associated with the first term and the two lateral ones associated with the third and seventh terms). The question we ask ourselves is how to extract morphological information about the sample from experimental measurement of the function $\tilde{r}_c(\tau,T)$ at a fixed delay $\tau=\tau_0$. In order to answer this question, we remark that the second and fifth terms in Eq.~(\ref{eq:espectrograma_qoct_aprox}) correspond to two peaks, centered at $\tau$ and at $\tau+T_s$, respectively. This implies that the separation between these two peaks directly yields $T_s$, i.e., the optical thickness of the sample. However, the existence of other peaks can complicate the correct identification of those two peaks, which bear useful information about the sample.

In this context, let us also note that the third, fourth, seventh and eighth terms in Eq.~(\ref{eq:espectrograma_qoct_aprox}) are proportional to $\mbox{cos}(\omega_0 T_s)$. This brings us to refer the reader to an earlier paper from our group \cite{Yepiz2019} in which we studied QOCT in the context of such a two-interface sample, in which we allowed the SPDC pump to be pulsed. In that paper, we show that terms such as these with the argument of the cosine function proportional to the pump frequency $2 \omega_0$, will tend to average out to zero as the pump bandwidth is allowed to increase, and can be entirely suppressed if the pump is in the form of a train of ultrashort (fs)  pulses. Therefore, for a sufficient pump bandwidth, the peaks associated with the cosine terms are suppressed, leaving only the central peak and the two peaks from which we can extract morphological information about the sample. It turns out that this can be generalized to any number of interfaces: by using a sufficient pump bandwidth the function $\tilde{r}_c(\tau,T)$ exhibits a central peak in addition to one peak per interface (appearing on both sides of the central peak), with the separation between peaks directly yielding the separation between interfaces in the sample. Also, the relative heights of the peaks directly yield information about the relative weights (determined by the reflectivities) of the contributions from each of the interfaces.

\begin{figure}[h!]
\centering
\includegraphics[width=6.5cm]{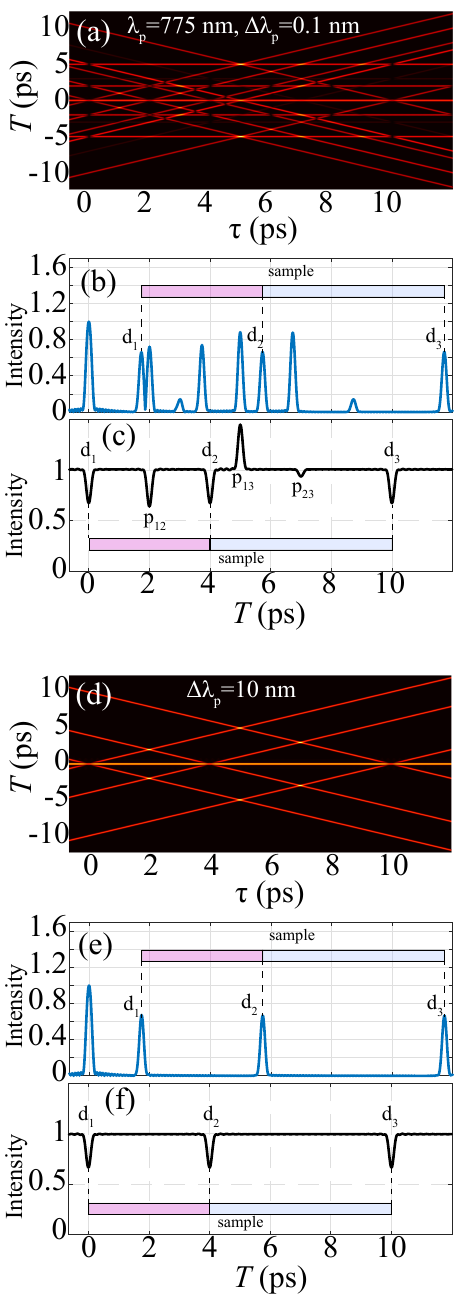}
\caption{(a) and (d) Simulation of the temporal-domain interferogram $\tilde{r}_c(\tau,T)$ for a 3-layer sample (intermediate layer at $40\%$ of the sample thickness, in addition to the two extremal interfaces); in (a) we show the case of an SPDC source centered at $775$~nm with a narrowband pump ($0.1$~nm), while in (d) we increase the pump bandwidth to $10$~nm.  (b) and (e) Evaluation of  $\tilde{r}_c(\tau,T)$ at  $\tau_0=-1.7$~ps; while (b) corresponds to a narrow pump bandwidth ($0.1$~nm), (e) shows the effect of increasing the bandwidth to $10$~nm. (c) and (f) HOM interferogram resulting for the above two cases; (c) for a narrow pump bandwidth ($0.1$~nm), and (f) for a pump bandwidth of $10$~nm.
 \label{fig:qoct_technique_3layer}}
\end{figure}

For further illustration of these ideas, let us consider a specific case involving three interfaces; see Fig.~\ref{fig:qoct_technique_3layer}. Specifically, we assume a sample that contains an intermediate interface at $40\%$ of the total sample thickness of $10$~ps in temporal units, as well as the two extremal interfaces. Figure~\ref{fig:qoct_technique_3layer}(a) shows a plot of the function $\tilde{r}_c(\tau,T)$ assuming a continuous-wave pump, while Fig.~\ref{fig:qoct_technique_3layer}(d) shows the same function resulting from a pump with a $\delta \lambda_p=10$ nm bandwidth. Note that the plot is significantly simplified, with fewer lines appearing as the pump bandwidth is increased. 

Let us now select a fixed delay given by $\tau_0=-1.7$~ps. Figure~\ref{fig:qoct_technique_3layer}(b) shows a plot of the resulting function $\tilde{r}_c(\tau_0,T)$ for a continuous-wave pump, where we have chosen not to display the peaks at negative values of $T$ on account of the symmetry in this function. In this case, there are $10$ peaks for the CW pump case, which are reduced to four peaks for a pulsed pump, a central one, and three additional peaks which directly yield morphological information about the sample.

Note that in the application towards QOCT of our frequency-resolved HOM measurement at a fixed delay $\tau=\tau_0$, it is not necessary for the determination of the desired morphological information to recover the standard delay-scanning HOM interferogram, i.e., we can directly recover this information from the function  $\tilde{r}_c(\tau_0,T)$ at a fixed delay $\tau_0$. However, this standard delay-scanning HOM interferogram \emph{can} be recovered, as we discuss below. We start by re-writing the expression for the HOM interferogram
\begin{equation}
R_{c}(\tau)=\frac{R_0}{2}\int_{-\infty}^{\infty} \mathrm{d} \Omega \abs{f(\Omega)}^2 \abs{H(\Omega)-H(-\Omega)e^{i\Omega \tau}}^2, \label{eq:Rc_tau_general}
\end{equation}
assuming that the function $f(\Omega)$ is symmetric, i.e.,  $f(-\Omega)= f(\Omega)$, for the case where one of the photons is reflected from a QOCT sample with a response function $H(\Omega)$ given by
\begin{equation}
H(\Omega)= \sum\limits_{j=0}^{N-1} r^{(j)} e^{i(\omega_0+\Omega) T_s^{(j)}} = r^{(0)}+r^{(1)}e^{i(\omega_0+\Omega)T_s^{(1)}}+\ldots,
\end{equation}
where $r^{(j)}$ is the reflectivity of the $j$th interface and $T_s^{(j)}$ is the time traveled in the round-trip from the $0$th layer to the $j$th layer (with $T_s^{(0)}$ chosen to be $0$, which corresponds to the positions of all interfaces measured with respect to the position of the first interface).

The HOM interferogram can then be recovered by following the following steps: 1) setting a fixed delay and experimentally acquiring the function $r_c(\tau_0,\Omega)$, 2) taking a numerical Fourier transform to obtain $\tilde{r}_c(\tau_0,T)$, 3) if the SPDC pump has a sufficient bandwidth, we take all peaks appearing on the right-hand side of the central peak and directly obtain the optical thicknesses $T_s^{(i)}$ from the left-most interface to the $i$th interface from the separation of the peaks, as well as the weight $r^{(i)}$ corresponding to each interface from the peak heights, thus constructing the function $H(\Omega)$, 4) numerically filter one peak, e.g., the central peak and take an inverse Fourier transform thus obtaining the function $f(\Omega)$, 5) numerically compute the HOM interferogram using Eq.~(\ref{eq:Rc_tau_general}). In Figs.~\ref{fig:qoct_technique_3layer}(c) and (f) we present the HOM dip recovered using this procedure for a CW pump and a pulsed (with $\Delta \lambda_p=10$ nm bandwidth) pump. Note that, as expected, for the pulsed pump case, the cross-interference intermediate structures can be fully suppressed.

%
%numerically filtering each of the peaks, 4) taking a numerical inverse Fourier transform for each filtered peak, 5) and multiplying the result by the correct phase derived from the Fourier shift theorem, allowing us to recover function $F(\Omega)$. In fact, in the symmetric case for which  $\vert f(-\Omega)\vert= \vert f(\Omega)\vert$, the function $F(\Omega)$ can be recovered from any one of the peaks.  With this function, we are now able to numerically compute the QOCT interferogram which would have been obtained in a standard delay-based HOM measurement, with the following equation which is equivalent to Eq. \ref{xxx} for the specific case of a symmetric joint amplitude,

% As is well known, in QOCT besides the dip corresponding to each interface, an intermediate structure which can be in the form of either a dip or a peak will appear for each pair of surfaces.  The presence of such intermediate structures can complicate the identification of HOM dips related to the actual interfaces in the sample.     In another recent paper from our group, we have demonstrated a quantum state engineering recipe which involves the use of an SPDC pump in the form of a train of ultrashort pulses, leading to the possibility of fully suppressing these intermediate structures\cite{xxx}.  We note that this technique could be successfully incorporated into our current HOM interferogram reconstruction protocol based on spectrally-resolved detection at a fixed temporal delay.

%%%%%%%%%%%%%%%%%%%%%%%%%%%%%%%%%%%%%%%%%%%%%%%%%%%%%%%%%%%%%%%%%%%%%%%%%%%%%%%%%%%%%%%%%%%%%%%%%%%%%%%%%%%%

\section{Experiment}

\begin{figure*}[h!]
\centering
\includegraphics[width=11cm]{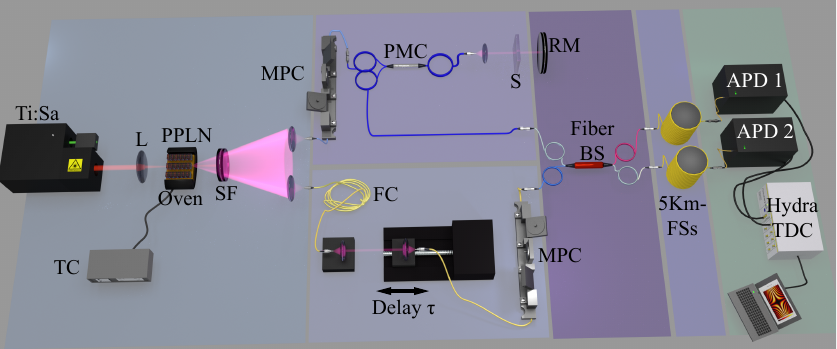}
\caption{Experimental setup. Ti:Sa: Titanium-Sapphire laser, TC: temperature controller, L: plano-convex spherical lens, PPLN: periodically-poled lithium niobate nonlinear crystal, SF: set of band-pass and long-pass filters, MPC: manual fiber polarization controller, PMC: polarization-maintaining optical circulator, FC: compensating fiber, S: sample, RM: reference
mirror, BS: beam splitter, FSs: fiber spools, TDC: time-to-digital converter, APD: avalanche photodetectors. \label{fig:setup}}
\end{figure*}

We have carried out an experiment in order to verify that frequency-resolved detection in a HOM interferometer, allows us to reconstruct the HOM dip without the need for varying the signal-idler temporal delay. Our experimental setup is depicted in Fig.~\ref{fig:setup}. Our SPDC source is based on a CW Ti:Sapphire laser, centered at $775$~nm, which pumps a periodically-poled lithium niobate crystal of $1$~cm thickness operated at a temperature of $90^\circ$~C, housed in a crystal oven with temperature precision of $\pm0.1^{\circ}$. The poling period ($\Lambda=19.1$~$\mu$m) is selected so as to permit frequency-degenerate, non-collinear SPDC at this temperature (with a $\pm 1.25^{\circ}$ propagation angle), thus producing photon pairs centered at $1550$~nm.

Both photons in a given pair are coupled into single-mode fibers, after being transmitted through a bandpass filter centered at $1550$~nm with $40$~nm 1/e full-width. In the case of the signal-photon arm, the fiber leads to one of the ports of a fiber circulator, so that this photon emanates into free space from a second port, is collimated with a lens (with $15$~mm focal length) and is reflected from a mirror (sample) for a HOM (QOCT) interferogram measurement so as to be re-coupled into the same port of the fiber circulator. The photon subsequently exits the circulator through the third port. In the case of the idler-photon arm, this photon is sent through a free-space delay; the photon is out-coupled using a lens with a $f=15$~mm focal length and coupled back into a single-mode fiber with an identical lens mounted, along with the fiber tip, on a computer-controlled translation stage (with minimum step of $200$~nm). The two photons then meet at a fiber-based beamsplitter (BS), with the two BS output ports each leading to a $5$~km spool of single-mode optical fiber, and from there to an InGaAs free-running avalanche photodiode. The electronic pulses produced by the APDs are sent to a Hydraharp time-to-digital converter so as to monitor, for each coincidence event, the signal and idler detection times with a $32$~ps resolution.

The signal and idler single-photon wavepackets propagating through the two fiber spools are temporally stretched since different frequencies travel at different group velocities. With adequate calibration, we are able to convert for each coincidence event the time of detection difference across the two output modes from the beam splitter into a measurement of the frequency detuning variable $\Omega$~\cite{Zielnicki2018,Yepiz2019}. Collecting data from multiple events, we build a histogram which corresponds to an experimental measurement of the joint spectral intensity for the signal and idler photons emerging from  the HOM beamsplitter output ports. Our experiment involves translating the free-space delay motor, and obtaining such a histogram at each delay stop, yielding an experimental measurement of the function $r_c(\tau,\Omega)$.

\begin{figure}[h!]
\centering
\includegraphics[width=8.5cm]{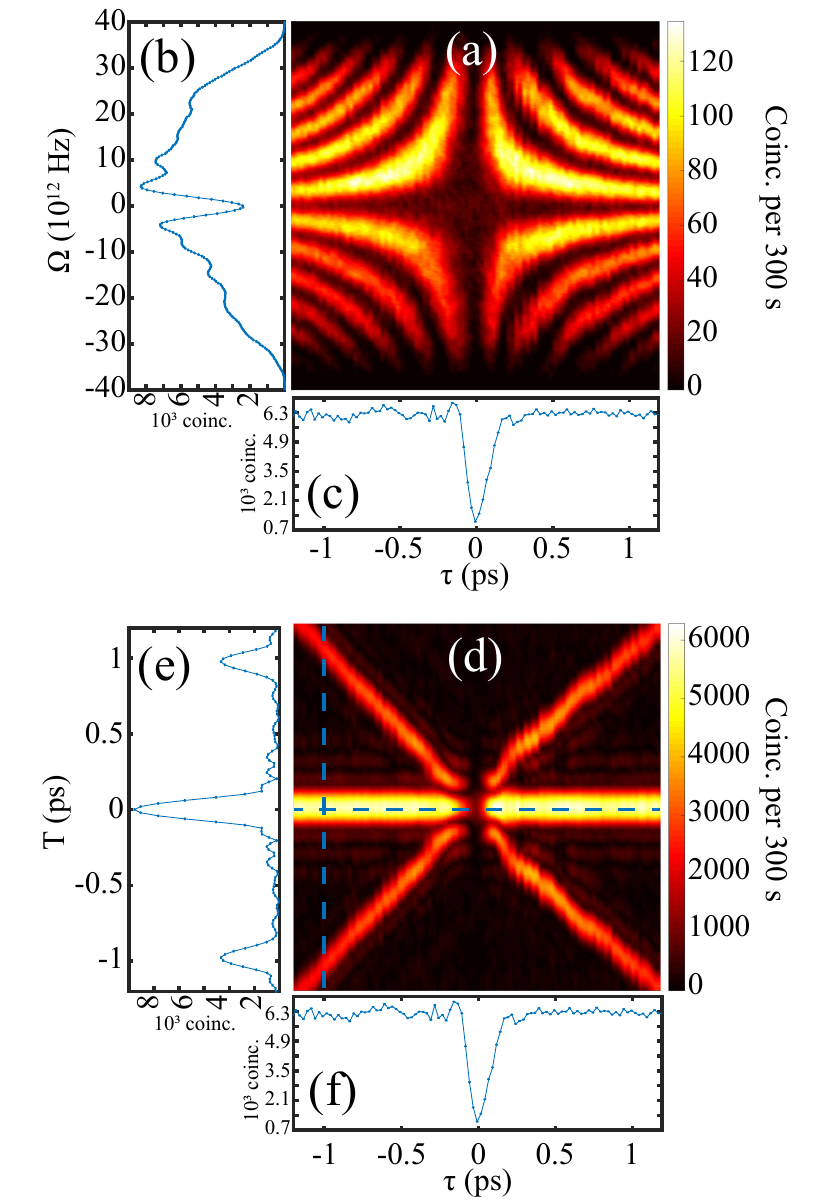}
\caption{(a) Experimental measurement of the delay-frequency interferogram $r_c(\tau,\Omega)$, for a single-layer sample (plain mirror). (b) Result of integrating the interferogram over $\tau$, yielding a HOM-like dip in the frequency variable $\Omega$. (c) Result of integrating the interferogram over $\Omega$, yielding the HOM interferogram.  (d) Numerical  Fourier transform of (a), yielding the time-domain interferogram $\tilde{r}_c(\tau,T)$. (e) Evaluation of $\tilde{r}(\tau,T)$ at $\tau=-1$~ps. (f) Evaluation of $\tilde{r}(\tau,T)$ at  $T=0$, yielding the HOM interferogram. 
 \label{fig:hom_espec_exp}}
\end{figure}

Figure~\ref{fig:hom_espec_exp}(a) shows an experimental measurement, thus obtained of the function $r_c(\tau,\Omega)$ for a single-interface sample in the form of a standard mirror, with a collection time of $300$~s per delay stop. As may be appreciated, we obtain an excellent agreement with the corresponding theoretical figure, see Fig. \ref{fig:hom_espec_calc}(a). In panel (c) of this figure, we show the standard HOM interferogram obtained from the measurement of $r_c(\tau,\Omega)$ by numerically integrating over the frequency $\Omega$. In panel (b) we show the effect of integrating the experimental data for $r_c(\tau,\Omega)$ over the delay $\tau$, which (as already discussed above), interestingly leads to a HOM dip-like structure vs. frequency instead of delay. In panel (d) we show the numerically-obtained Fourier transform of $r_c(\tau,\Omega)$, i.e., the function $\tilde{r}_c(\tau,T)$. As may be appreciated, we likewise observe an excellent agreement with the corresponding theoretical figure (see Fig.~\ref{fig:hom_espec_calc}).

\begin{figure}[h!]
\centering
\includegraphics[width=8.5 cm]{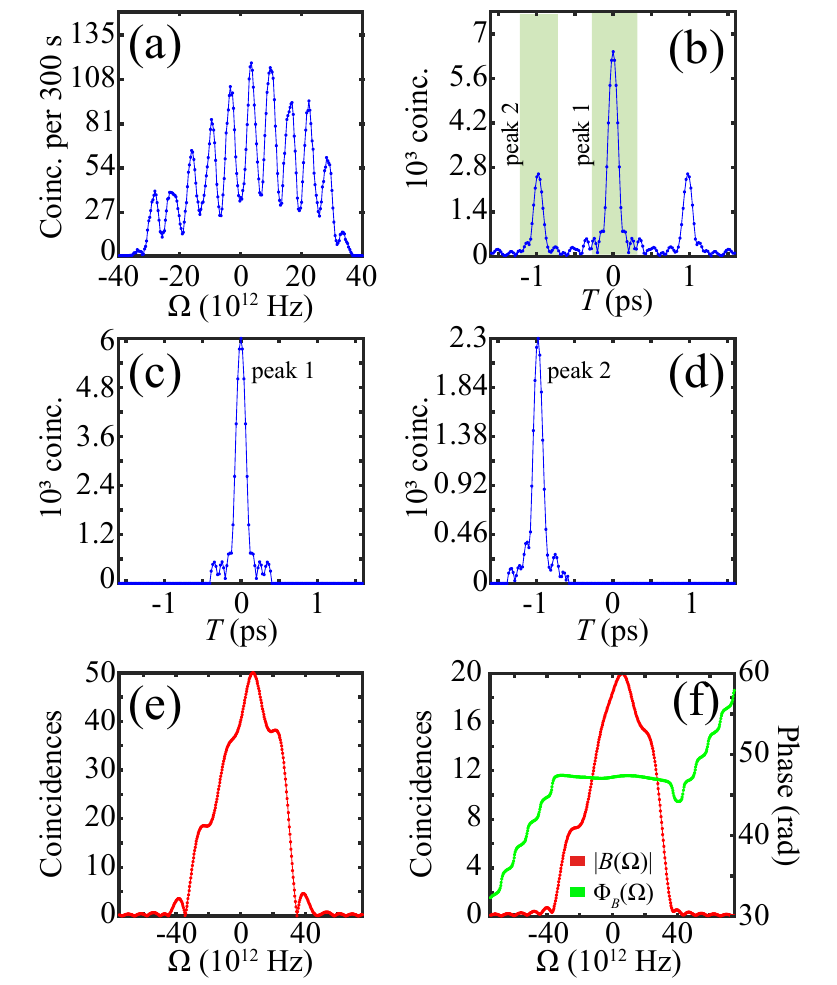}
\caption{Reconstruction procedure for the functions $A(\Omega)$ and $B(\Omega)$, from which we can compute the HOM interferogram through Eq.~(\ref{eq:rc_HOM_develoment}). (a) Evaluation of the delay-frequency interferogram $r_c(\tau,\Omega)$ at $\tau=-1$~ps. (b) Numerical Fourier transform of (a), yielding $\tilde{r}_c(\tau_0,T)$ with $\tau_0=-1$~ps. (c) and (d) Peaks $1$ and $2$ isolated from $\tilde{r}_c(\tau_0,T)$ by restricting the $T$ variable to the two windows indicated in panel (b). (e) Function $A(\Omega)$ obtained as the inverse Fourier transform of peak $1$. (f) Function $B(\Omega)$ obtained as the inverse Fourier transform of peak $2$, multiplied by the phase $\exp(i \Omega \tau_0)$; both amplitude and phase are shown. \label{fig:algoritmo_recuperacion}}
\end{figure}

In Fig.~\ref{fig:algoritmo_recuperacion} we outline the protocol used for the reconstruction of the HOM interferogram, relying on frequency-resolved coincidence detection of the HOM beamsplitter output modes, at a fixed delay $\tau=\tau_0$. For the same experimental situation corresponding to Fig.~\ref{fig:hom_espec_exp}(a), we have selected a fixed delay $\tau_0=-1.018$~ps, effectively obtaining a vertical `slice' of the plot in Fig.~\ref{fig:hom_espec_exp}(a). The resulting interferogram $r_c(\tau_0,\Omega)$ at this fixed delay is displayed in Fig.~\ref{fig:algoritmo_recuperacion}(a). In panel (b) we show the numerical Fourier transform of $r_c(\tau_0,\Omega)$, i.e., $\tilde{r}_c(\tau_0,T)$, along with two temporal windows which encompass the central peak (labeled peak $1$) and the left-hand peak (labeled peak $2$). Panel (c) shows peak $1$ filtered out from $\tilde{r}_c(\tau_0,T)$, while panel (e) shows the numerical inverse Fourier transform of this filtered peak, corresponding to our estimation of function $A(\Omega)$. Panel (d) shows peak $2$, filtered out from $\tilde{r}_c(\tau_0,T)$, while panel (f) shows the numerical inverse Fourier transform of this filtered peak multiplied by $\mbox{exp}(i \tau \Omega)$, corresponding to our estimate for the function $B(\Omega)$; we have shown both the absolute value and the phase. Note that while from our theory, the function $A(\Omega)$ is expected to be symmetric as already mentioned above, in the experimental measurement we obtain a slight asymmetry due to experimental imperfections, which is also evident in the HOM dip (see below).

\begin{figure}[h!]
\centering
\includegraphics[width=8 cm]{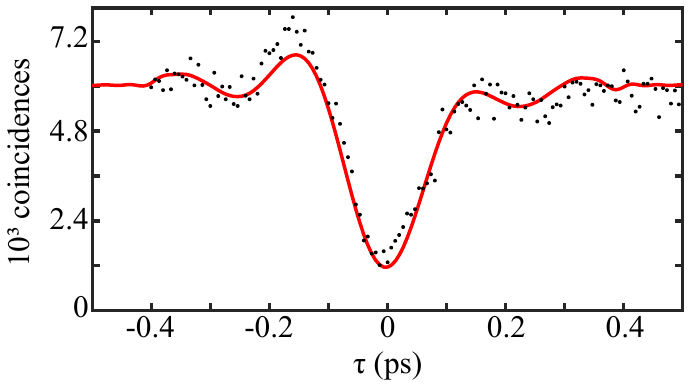}
\caption{Reconstructed HOM dip  (red line) and conventional HOM dip obtained through scanning the delay with non-frequency resolved coincidence counting (black dots).
\label{fig:hom_recuperado}}
\end{figure}

Once we have numerical estimates for the functions $A(\Omega)$ and $B(\Omega)$ obtained from our experimental measurement of $r_c(\tau_0,\Omega)$, we are in a position to recover the HOM dip through numerical integration of Eq.~(\ref{eq:rc_HOM_develoment}); the result is shown in Fig.~\ref{fig:hom_recuperado} (red continuous line). For comparison purposes, we have also carried out a standard, non-frequency-resolved HOM measurement by monitoring the coincidence counts vs. delay, with an acquisition time of $300$~s per delay stop. The result of this measurement is also shown in Fig.~\ref{fig:hom_recuperado} (black dots). As is clear, we obtain an excellent agreement between the recovered HOM dip obtained through frequency-resolved detection at a fixed delay, and the directly-obtained standard HOM dip.    Note the visibility does not reach 100\% because of slight asymmetries in the joint spectral intensity; in our experiment the visibility reaches close to 100\%  when filtering the photon pairs with a narrowband pass filter.

\begin{figure}[h!]
\centering
\includegraphics[width=8cm]{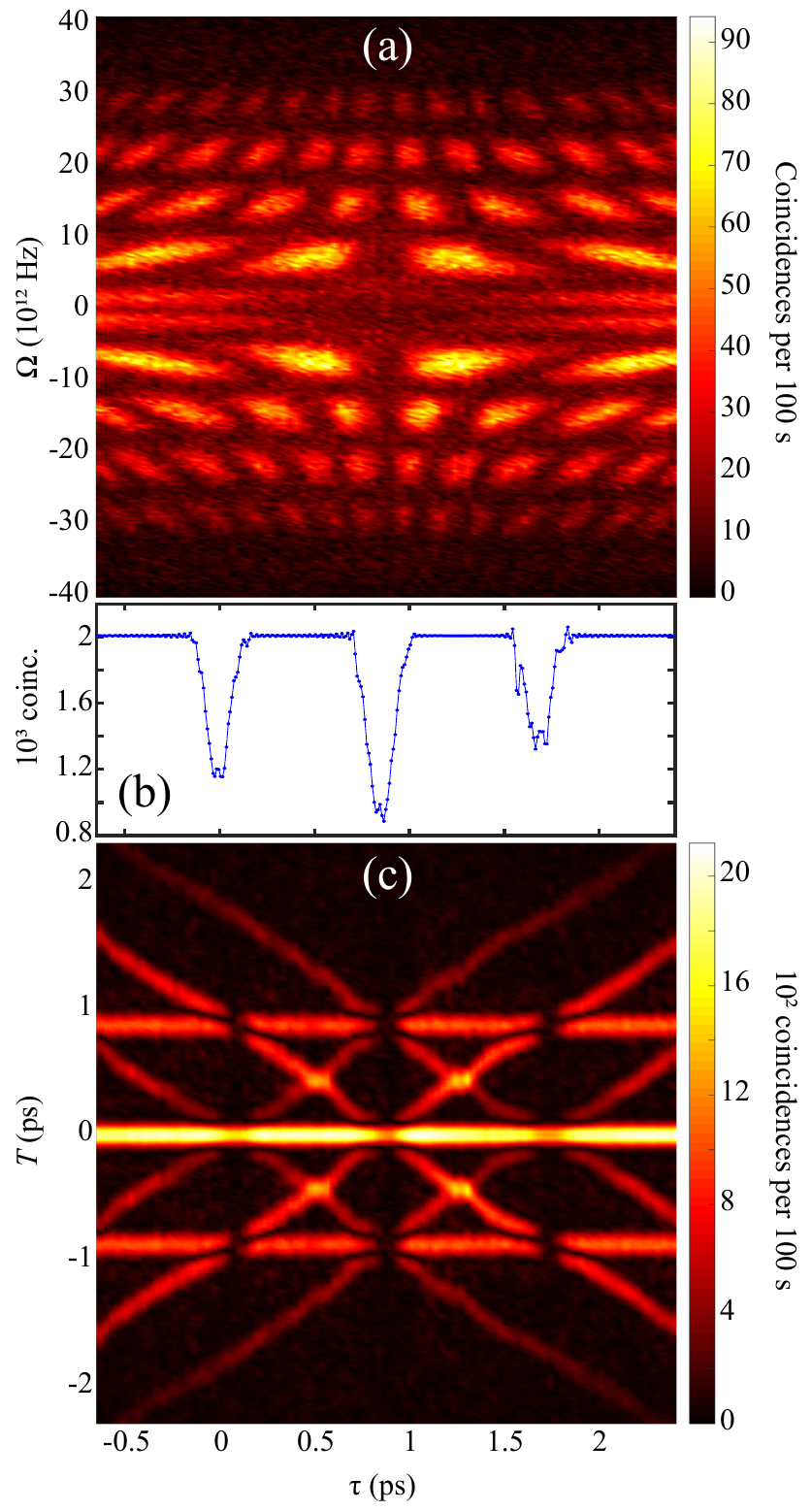}
\caption{a) Experimental measurement of the delay-frequency interferogram $r_c(\tau,\Omega)$, for a two-layer sample (borosilicate glass coverslip of $170$~$\mu$m thickness). (b) Result of integrating the interferogram over $\Omega$, yielding the QOCT interferogram. (c) Numerical Fourier transform of (a), yielding the time-domain interferogram $\tilde{r}_c(\tau,T)$. \label{fig:qoct_espec_exp}}
\end{figure}

In order to illustrate these ideas, as applied to QOCT, we have repeated the experiment above in such a way that one of the photons is reflected from a two-interface sample instead of from a mirror. The sample used is a borosilicate glass coverslip of $170$~$\mu$m thickness. In Fig.~\ref{fig:qoct_espec_exp}(a), we show an experimental measurement thus obtained of the function $r_c(\tau,\Omega)$ with an acquisition time of $100$~s per delay stop. We also show, in Fig.~\ref{fig:qoct_espec_exp}(b), the function $\tilde{r}_c(\tau,T)$ obtained as the numerical Fourier transform of the experimental data for function $r_c(\tau,\Omega)$. In panel (c) we show the standard delay-based HOM interferogram obtained as the numerical integration of the experimentally-obtained function $r_c(\tau,\Omega)$ over $\Omega$. Note that all of these experimental plots exhibit an excellent agreement with the corresponding theory plots, shown in Fig.~\ref{fig:qoct_espec_calc}.

\begin{figure}[h!]
\centering
\includegraphics[width=8cm]{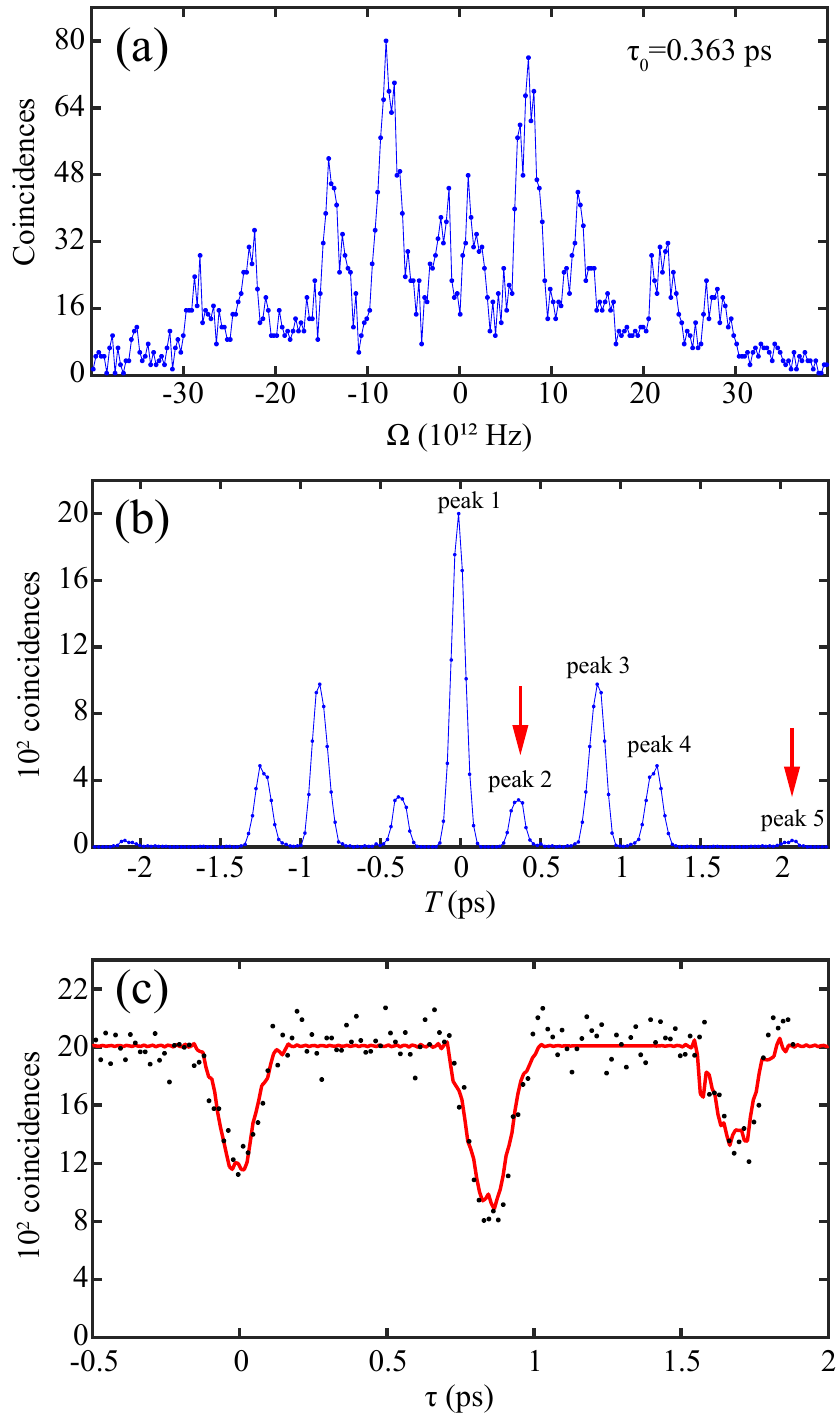}
\caption{Reconstruction of the sample morphology and QOCT interferogram for a two-layer sample (borosilicate glass coverslip of $170$~$\mu$m thickness). (a) Experimental measurement of the function $r_c(\tau_0,\Omega)$ at a fixed delay $\tau_0=0.363$~ps, (b) Numerical Fourier transform of (a), yielding $\tilde{r}_c(\tau_0,T)$; here we have labeled 5 of the resulting peaks with the numbers $1$-$5$. (c) Reconstructed QOCT interferogram (red line) and conventional  delay-scanning, non-spectrally-resolved HOM measurement (black points).
 \label{fig:qoct_recuperacion}}
\end{figure}

In Fig.~\ref{fig:qoct_recuperacion} we summarize the reconstruction of i) the morphological information of the sample, and ii) the expected delay-scanning HOM interferogram, from our frequency-resolved HOM measurement at a fixed delay. We select a fixed delay $\tau_0=0.363$~ps, effectively obtaining a vertical `slice' of the plot in Fig.~\ref{fig:qoct_espec_exp}(a) corresponding to the function $r_c(\tau_0,\Omega)$, which has been plotted in Fig.~\ref{fig:qoct_recuperacion}(a). In panel (b) we show the numerical Fourier transform of the function $r_c(\tau_0,\Omega)$, i.e., thus obtaining the function $\tilde{r}_c(\tau_0,T)$. This function exhibits the nine peaks predicted by Eq.~(\ref{eq:espectrograma_qoct_aprox}). While our experiment was carried out with a CW pump for the SPDC process, and therefore we cannot eliminate all peaks with amplitudes proportional to $\mbox{cos}(\omega_0 T_s)$, we have labeled with red arrows the two terms (second and fifth in Eq.~(\ref{eq:espectrograma_qoct_aprox})) which bear morphological information about the sample. We can directly obtain the optical sample thickness $T_s$, as well as the weights $r^{(0)}$ and  $r^{(1)}$, from the separation and heights of these two peaks. Following the recipe outlined above for the reconstruction of the HOM dip, in Fig.~\ref{fig:qoct_recuperacion}(c) we show the result of such a reconstruction, along with a direct, delay-scanning measurement (the latter with an acquisition time of $100$s per delay stop). Clearly, there is an excellent agreement between the two measurements.

We note that a frequency-domain HOM effect was demonstrated by Kobayashi \emph{et al.}~\cite{Kobayashi2016}.  However, we clarify that their scheme is fundamentally different from ours in two aspects. First, they use a frequency converter to realize destructive quantum interference between indistinguishable frequency modes, instead of spatial modes. Second, they do not spectrally-resolve the output state, which is the central aspect of our work.

We also note that the original QOCT technique is analogous to time-domain OCT, which is typically associated with the term OCT. However, variations on OCT include full-domain OCT (whose quantum version appears on Ref.~\cite{Ibarra2020}), swept-source OCT~\cite{Chinn_97}, and spectral-domain OCT (SD-OCT)~\cite{Morgner_00}. The latter uses a spectrometer to analyze the resulting interference pattern as a function of wavelength while the reference arm is stationary. Thus we have demonstrated, for the first time to the best of our knowledge, the quantum version of SD-OCT, which could be referred to as spectral-domain QOCT. We hope that our work will facilitate the implementation of practical QOCT devices and inspire quantum-mimetic experiments~\cite{Erkmen2006,Kaltenbaek2008,Kaltenbaek2009}.

%%%%%%%%%%%%%%%%%%%%%%%%%%%%%%%%%%%%%%%%%%%%%%%%%%%%%%%%%%%%%%%%%%%%%%%%%%%%%%%%%%%%%%%%%%%%%%%%%%%%%%%%%%%%%%%%%%%%%%%%%%%%%%%%%%%%%%%%%%%%%%%%%%%%%%%%%%%
\section{Conclusions}

In this paper, we have studied HOM interferometry from a new perspective, i.e., by allowing spectral resolution of the single-photon detectors, which, as we show, permits the recovery of the HOM dip without the need for varying the delay between the incoming signal and idler photons. Concretely, we have shown, both from theoretical and experimental standpoints, that setting the delay to a fixed value (greater than the dip half-width) and by enabling spectrally-resolved coincidence photon counting, we can recover the HOM interferogram with the same level of counts that would have been obtained through a standard, non-spectrally-resolved HOM measurement (involving a sufficient number of delay stops for an adequate sampling of the dip).  We have also shown that our technique allows for the reconstruction of the symmetrized spectral intensity or the standard HOM interferogram, depending on the two-photon state symmetry properties, with full dispersion cancellation.

We have presented experimental measurements, along with simulations, for the spectral-delay HOM interferogram in two different cases: a single-interface sample (i.e., a plain mirror) and a two-interface sample. From the data at a single delay value, for each of these two cases,  we have recovered through the procedure presented here the HOM interferogram and have compared it with a standard HOM measurement based on delay scanning and non-spectrally-resolved coincidence counting, exhibiting an excellent agreement.   We have also presented a simulation concerning a three-layer sample so as to illustrate the application of our technique for QOCT in the context of a more general sample.  The importance of these results is that the time required in order to acquire a HOM interferogram can be drastically reduced since a single delay stop is required. This is expected to be of particular importance in the context of QOCT, for which delay scanning results in long acquisition times, which is impractical in real-life clinical settings.

\section*{Funding Information}
PAPIIT (UNAM) (IN104418); AFOSR (FA9550-16-1-1458); Consejo Nacional de Ciencia y Tecnolog\'{i}a (CONACYT) (Fronteras de la Ciencia 1667, 293471, 295239, APN2016-3140).

% PAPIIT (UNAM) grant IN104418; CONACYT Fronteras de la Ciencia grant 1667; AFOSR grant FA9550-16-1-1458.

%\section*{Acknowledgments}
%Formal funding declarations should not be included in the acknowledgments but in a Funding Information section as shown above. The acknowledgments may contain information that is not related to funding:
%The authors thank H. Haase, C. Wiede, and J. Gabler for technical support.

\section*{Disclosures}
The authors declare no conflicts of interest.

% Bibliography
\bibliography{biblio}
\end{document}